# Workbook Structure Analysis – "Coping with the Imperfect"

Bill Bekenn and Ray Hooper
Fairway Associates Ltd. PO Box 846, Ipswich IP5 3TT UK
info@fairwayassociates.co.uk

**ABSTRACT**

*This Paper summarises the operation of software developed for the analysis of workbook structure. This comprises: the identification of layout in terms of filled areas formed into "Stripes", the identification of all the Formula Blocks/Cells and the identification of Data Blocks/Cells referenced by those formulas. This development forms part of our FormulaDataSleuth® toolset. It is essential for the initial "Watching" of an existing workbook and enables the workbook to be subsequently managed and protected from damage.*

## 1  INTRODUCTION

A study [Lawson et al, 2007] has shown that inexperienced spreadsheet developers have markedly less understanding of the benefits of practices such as integrating spreadsheet modules, keeping data and formulas separate and testing than experienced developers. This can lead to critical spreadsheets and workbooks being vulnerable to error because of poor workbook structure. This is a particular problem in small to medium sized organisations where resources to support spreadsheet development are limited. We have developed a software algorithm to enable the structure of a Workbook to be analysed and subsequently improved.

There is a school of thought that there is no such thing as "spreadsheet good practice" [Colver, 2004], arguing that there are many different ways of designing a workbook, each of which has advantages and disadvantages. In contrast, our experience of spreadsheet design and use shows that there is "good practice" at least for spreadsheet layout. This "good practice" is often adopted intuitively by experienced spreadsheet practitioners. This is distinct from overall workbook design (e.g. having Inputs, Outputs and Workings in separate sheets). Layout good practice can be based on assessing whether modifications (e.g. inserting rows) are likely to cause damage and tracking whether Blocks and Single Cells are accurately referenced by formulas.

Algorithms have been developed that perform Workbook Structure Analysis and find all the Formula Blocks/Cells and Data Blocks/Cells. To achieve this all the Formula Blocks are examined and the ones that share a common width or height are identified. The relationships between the Formula Blocks/Cells can thus be recorded. The Data Blocks/Cells are then found from the Formula References. Manual inspection by an expert practitioner could achieve the same end. These algorithms have evolved from a "Find Formulas" procedure that originally found formulas only and displayed the structure as borders around the filled Formula Blocks. This feature is not unique and is included in products such as OAK V4 [OPERIS, 2009] and SpACE [AuditWare, 2006].

Why perform the complex analysis needed to reveal the workbook layout and structure beyond identifying filled Blocks? At least one recent study [Aurigemma, 2010] has shown that automated analysis performs less well at finding errors in spreadsheets than detailed study by a person. Our Workbook Structure Analysis does not find all the errors and instead concentrates on what can be done automatically to reveal the layout. It then checks whether references in formulas accurately follow the layout/structure. Once the underlying structure is revealed then areas of a workbook that appear untidy, e.g. with Blocks that conflict with the overall structure, can be targeted for manual detailed analysis. Our experience of using the algorithm, on real spreadsheets from our client base, is that revealing the structure greatly speeds up the process of understanding and analysing a workbook. This addresses some of the problems encountered when needing to understand legacy spreadsheets [Hodnigg, 2008]. As well as visually revealing the structure, potentially incorrect cell references are indicated as warnings or errors for the user to inspect. The algorithm also gives an early indication of the likely scope of the problems as a percentage "score", and will show when the workbook is past redemption and needs to be re-worked rather than checked [Murphy, 2007].

The original version of the "Find Formulas" procedure did not record spreadsheet structure which could only be determined by the user inspecting the borders. Nevertheless, familiar patterns reflecting both good practice and poor layouts could be observed. Experience gained from manually analysing/checking spreadsheets has now enabled the enhancement of our *F*ormula*D*ata*S*leuth[®] toolset [Bekenn, 2008]. The latest version of the "Find Formulas and Data" procedure records the spreadsheet structure and flags layout anomalies. The correct identification of spreadsheet structure enables the accurate detection of all Data Blocks/Cells. This includes identifying, if present, Data Blocks which may be "unknown" to the user/developer but are nevertheless referred to by Formula Blocks. These are denoted "Unintentional Data Blocks".

## 2    STRUCTURE, "STRIPES", AND "BLOCKS"

Spreadsheet structure may be defined in terms of "Stripes" and "Blocks" which are defined in figure 1 and the following sub-sections. These concepts are examined in more detail in [Bekenn, 2009].

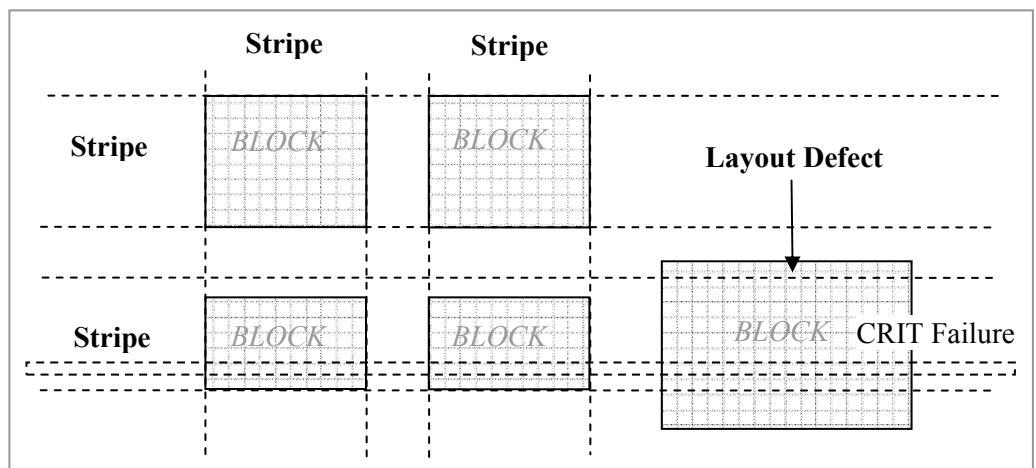

Figure 1: Spreadsheet Structure

### 2.1   "Blocks" and "Single Cells"

An area of data or filled formulas is referred to as a "Block" [O'Beirne, 2005]. Blocks are considered distinct from "Single Cells" which contain isolated items of data or one-off formulas. There are three types of Block:-

1. Single Row,
2. Single Column,
3. Two Dimensional (Multiple Row and Column).

## 2.2 "Stripes"

A horizontal set of Blocks spanning the same rows, make up a "Horizontal Stripe". A vertical set of Blocks spanning the same columns, make up a "Vertical Stripe". Stripes are more important if they contain multiple rows/columns than if they contain just single rows/columns.

## 2.3 "Idealised Layout"

An "Idealised Layout" is a layout composed only of Stripes. These stripes can have whole rows or whole columns inserted at any point, without introducing the risk of an error. This is indicated in figure 1 with the Vertical Stripes and the top Horizontal Stripe. Stripes are not allowed to overlap and must occupy their own set of rows or columns. A Vertical Stripe can intersect a Horizontal Stripe and thus a Two Dimensional Block can be part of both. Where possible, references within formulas should be to cells on the same horizontal row or the same vertical column.

It is possible to conceive a Column or Row Insertion Test ("CRIT") to test the resilience of a spreadsheet to modification. Any contiguous assembly of blocks which cannot have whole rows/columns inserted without causing errors is considered as failing CRIT.
An example of this situation is shown in figure 1 with the block at the bottom right. An "Idealised Layout" will always pass a CRIT.

## 3 HOW THE STRUCTURE ANALYSIS WORKS

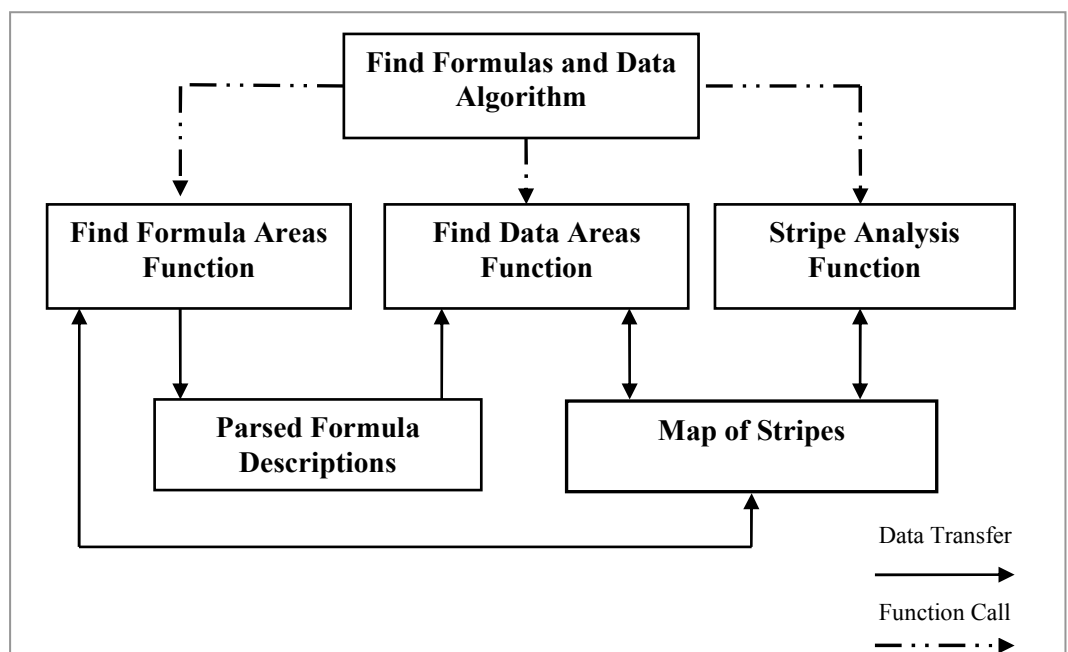

Figure 2: Program Components

Figure 2 shows the main program components required to execute the Workbook Structure Analysis. The Find Formulas and Data algorithm calls three functions to (i) find the Formula Blocks, (ii) find the Data Blocks and (iii) perform the Stripe Analysis. Any isolated formulas and data in Single Cells are also found. The functions interact with the two data sets containing the "Parsed Formula Descriptions" and the "Map of Stripes". Using these components a workbook structural analysis is performed in broadly the following order:-

1. Formula Blocks are found and logged in the "Map of Stripes" using the Find Formula Areas Function.
2. The Formula Blocks are parsed and recorded in the "Parsed Formula Descriptions" also using the Find Formula Areas Function.
3. Using the "Parsed Formula Descriptions" data set the Data Blocks are found and logged in the "Map of Stripes".
4. The Stripe Analysis function is invoked after 1, 2 and 3, to progressively refine the layout picture of the workbook contained in the "Map of Stripes".

The algorithm works like an expert practitioner to first record all major Blocks and their interrelation and then move on to more detailed checking. The detailed checking is capable of identifying Blocks that do not fit well with the overall structure, i.e. Blocks that do not conform to an "Idealised Layout".

Section 4 will describe in more detail the three functions and their interaction with the data sets. Section 5 will then explain in more detail how the algorithm uses the three functions described in section 4.

### 4   THE FIND AND STRIPE ANALYSIS FUNCTIONS

Three forms of information are recorded as the three functions are invoked:-

1. The location (cell address) of each Formula or Data Block/Cell, stored together with the "Parsed Formula Descriptions".
2. The locations (cell addresses) referenced within each formula, stored within the "Parsed Formula Descriptions".
3. The locations (row/column addresses) of the stripes detected by the functions, stored within the "Map of Stripes".

### 4.1   The "Find Formula Areas" Function

This function finds the filled Formula Blocks. A Formula Block would be expected to contain a contiguous area of filled formulas that could be generated from a single cell instance (seed) within the area by "Fill" or Copy and Paste. This is a relatively simple matter for the function to achieve with a contiguous area as it only needs to find the extent of the "Fill". A more complex situation is depicted in figures 3 & 4, where the extent of the "Fill" can be seen but the area is not contiguous, i.e. there are "Exceptions" within the area in the form of different formulas, data cells and blanks. These are shown in grey cross-hatch in figures 3 & 4.

There are two options which can be specified when this function is called, Simple Find (figure 3, any Exceptions will bound Blocks) and Complex Find (figure 4, defining the maximum extent of the Formula Blocks while skipping over any Exceptions).

As formula Bocks and Cells are found they are recorded in the "Map of Stripes" so as to be ignored if they are found by subsequent passes of the algorithm.

### 4.1.1 Simple Find (Exceptions bound Blocks)

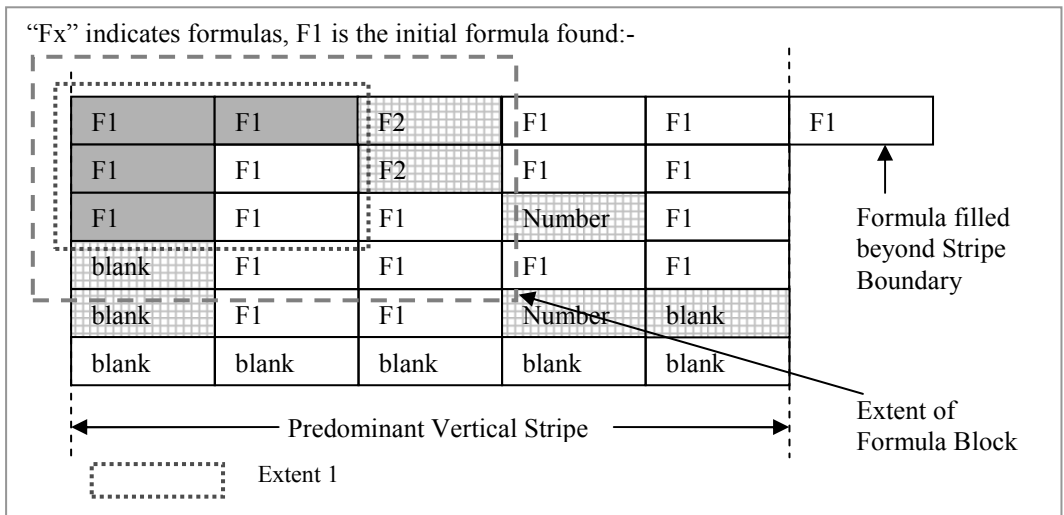

Figure 3: Finding a Formula Block with Simple Find

The approximate sequence of operations, used for establishing the initial layout picture, is as follows:-

1. Find a new, un-recorded, formula (scanning from top right of a sheet to bottom left across columns and then down rows).
2. Assuming the new formula is in the top left cell of a Block, scan across the top row and down the left column, looking for the first cell containing something different from the top left cell. These cells, shown in grey, are used to define the potential extent of a simple filled Block (figure 3, Extent 1).

The dotted outline in figure 3 indicates Extent 1 for the simple formula Block containing F1. Initially, this extends to two cells across and three down. There is the possibility that other rows and/or columns may be filled to a greater extent.

3. Rows to the right of Extent 1 and columns below it are checked to see if some rows/columns have a greater extent for the fill. If so, the size of the filled Block is increased to match the greatest horizontal and vertical extent finally defining the filled Block.

In figure 3 the top left filled Block found by the sequence of operations above, shown by the grey dashed outline "Extent of Formula Block" in figure 3, is logged in the "Map of Stripes". Other filled Blocks will also be found and logged.

### 4.1.2 Complex Find (Exceptions skipped)

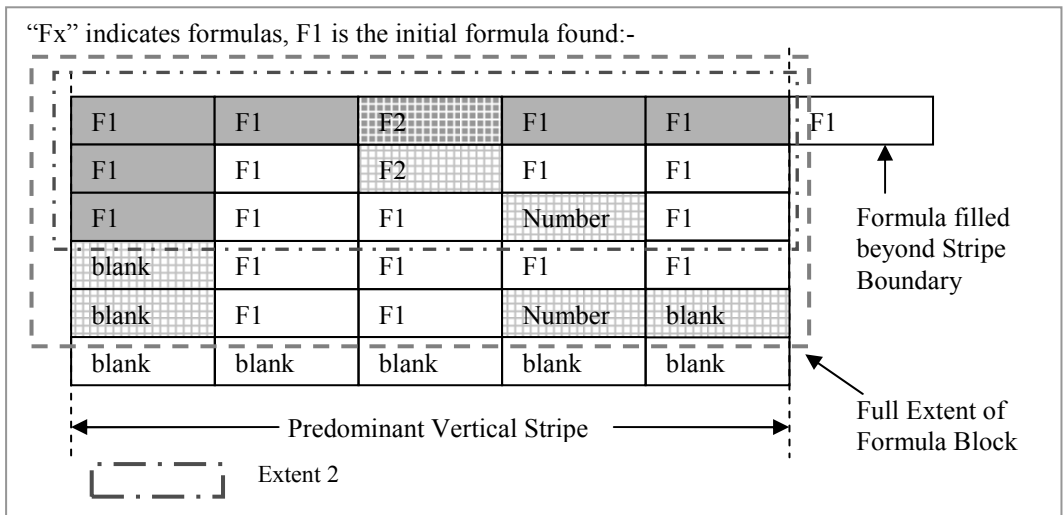

Figure 4: Finding a Formula Block with Complex Find

The approximate sequence of operations, used for finding Formula Blocks, is as follows:-

1. Find a new, un-recorded, formula (scanning from top right of a sheet to bottom left across columns and then down rows).
2. Assuming the new formula is in the top left cell of a Block, scan across the top row and down the left column looking for the end of a "Region" which is either delimited by a blank cell, a cell containing text, an existing predominant Stripe boundary or a formula that has already been found.
3. Scan back across the top row and up the left column from the end of the Region to find the furthest extent of formulas matching the one found initially in 1. These cells, shown in grey, are used to define the potential extent of a complex filled Block (figure 4, Extent 2).
4. Continue the backward scan, checking the top row and left column of Extent 2 to see how many cells contain either formulas that differ from the one found initially or numeric data.
5. If the number of different cells (separately across and down) is large compared with the number of identical ones then it is assumed that the filled Block is smaller than Extent 2in 3. In this case the scan is repeated forwards, looking for the first different cells, across and down, which are used to define the extent of a smaller filled Block.

The chained outline in figure 4 indicates Extent 2 for the complex formula Block containing F1. Initially, this extends to five cells across and three down. There is the possibility that other rows and/or columns may be filled to a greater extent. There is also no certainty that all the cells within the Block have the same formula.

6. Rows to the right of Extent 2 and columns below it are checked to see if some rows/columns have a greater extent for the fill. If so, the size of the filled Block is increased to match the greatest horizontal and vertical extent finally defining the filled Block.
7. Operations similar to 1 through 6 are repeated within the extent of the defined filled Block. These look for filled Blocks and Single Cells containing a different formula and for cells overwritten with data. All of these are marked as "Exceptions" and are potential errors.

The filled Block found by the sequence of operations above, shown by the grey dashed outline "Full Extent of Formula Block" in figure 4, is logged in the "Map of Stripes".

**4.2    The "Find Data Areas" Function**

Each formula in a workbook is analysed (or parsed) and recorded in the Parsed Formula Description data set (figure 2). This is done for each Formula Block as they are found by the Find Formula Areas function (section 4.1 and 4.1.2). The locations of the top left and bottom right cells of each range referenced by the formula are also recorded. A filled reference, where the row and/or column are relative, is taken as referring to a range of the same size as the filled Formula Block.

Data Blocks are found in stages using the different types of reference in the order of precedence below:-

1. Data referenced by relative filled formula References,
2. Data referenced by Range References but not overlapping with 1,
3. Data referenced by Single Cell References but not included in 1 or 2,
4. Data referenced by Range References but falling outside or in between Blocks/Cells that have been identified in 1, 2 or 3.

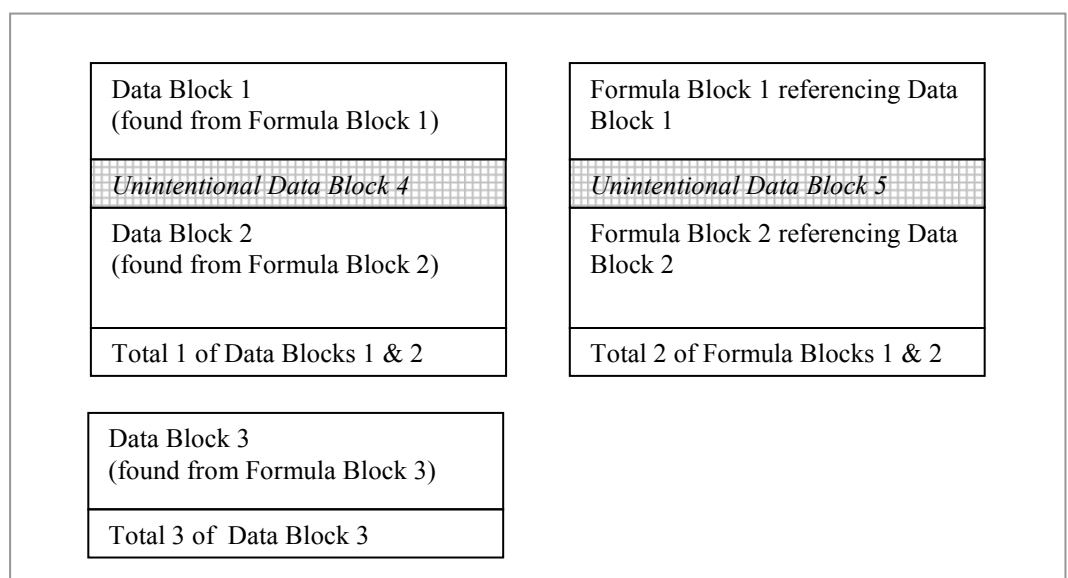

Figure 5: Finding Data Blocks

Figure 5 shows a common sheet layout where Data Blocks 1 & 2, separated by a blank row, are summed together into Total 1. Similarly Formula Blocks 1 & 2 are summed together into Total 2. The blank cells referenced are shown in grey cross-hatch and designated as "Unintentional Data Blocks". Formula Blocks 1 & 2 reference Data Blocks 1 & 2 respectively.

In figure 5 there are two types of Formula Block. Firstly there are the SUM() functions in Totals 1, 2 and 3. These will have range references. Secondly there are filled references in Formula Blocks 1 and 2. Which of these Formula Block types should be dealt with first?

It turns out that, from a structure analysis point of view, the order of precedence for the different types of reference is important. If range references were considered first then the Unintentional Data Block 4 would be combined with Data Blocks 1 & 2. This would form a single combined Data Block, which is not a correct analysis of the structure. Moreover, it conflicts with the Horizontal Stripes defined by Formula Blocks 1 & 2. On the other hand, if the filled references in Formula Blocks 1 & 2 are considered first, then the Data Blocks 1 & 2 are correctly identified. The Unintentional Data Blocks 4 & 5, found from the range references in Totals 1 & 2, are treated as separate. Data Block 3 is also found when considering the range reference in Total 3, as there is no filled reference to consider in this case. The "rule" is therefore to consider filled references before range references.

Each data block found is logged in the "Map of Stripes". This map is processed by the Stripe Analysis function (section 4.3), which identifies the ones define the sheet structure.

### 4.3 The "Stripe Analysis" Function

The Stripe Analysis function processes the Map of Stripes (see figure 2) to increasingly make sense of what the Find Formulas and Find Data functions discover from each pass of the overall algorithm. The map initially contains a mixture of (i) stripes which define the sheet layout picture, (ii) stripes that exist due to anomalies in fill and (iii) stripes caused by Exceptions within the Formula Blocks found. The number of filled rows and columns are totalled for each of the Vertical and Horizontal Stripes recorded in the map. These counts are then used to determine an order of predominance for each Stripe with the highest counts coming first.

Where a workbook conforms strictly to the criteria for an Idealised Layout (section 2.3) all the Stripes logged in the "Map of Stripes" could be used to define the structure of each sheet. However, the Idealised Layout criteria are not often met. It is necessary, therefore, to identify predominant Stripes within the map so that any layout anomalies do not interfere with the analysis of the sheet structure.

In general, therefore, the "Map of Stripes" is accessed starting with those stripes with the highest predominance. These are found first and are used to define the structure of each sheet. Any other stripes overlapping with the ones with highest predominance are noted as not conforming to the Idealised Layout criteria. However, any single column or row stripe that overlaps with a predominant Vertical or Horizontal Stripe will conform to an Idealised Layout. This is because single column or row Stripes will not cause CRIT failures.

### 5 THE "FIND FORMULAS AND DATA" ALGORITHM

The "Find Formulas and Data" algorithm is a multi pass procedure. It progressively builds up a record of the predominant Formula Blocks that obey Idealised Layout criteria (i.e. form "Stripes"). It then identifies other Formula Blocks and Cells that do not obey the Idealised Layout criteria. After each of these passes the Data Blocks are found using the different types of reference as described in section 4.2.

The concept of an Idealised Layout is important as the Blocks within a spreadsheet that conform to the rules for "Stripes" (section 2.2) take precedence over Blocks that do not. There are therefore two sets of passes with the first set tackling only Blocks that obey the Idealised Layout criteria and the second set tackling the rest.

Figure 6 shows a simple spreadsheet layout where Totals 1 & 2 are filled to 5 columns wide to include the row Totals 3 & 4. Formula Blocks 1, 2 & 3 are referencing Data Blocks 1, 2 & 3 respectively using relative references. Whilst Totals 1 & 2 are single filled Blocks, the right hand column of the Totals rows (shown with diagonal hatching) have a logically different function from the other totals and should be treated separately in terms of layout.

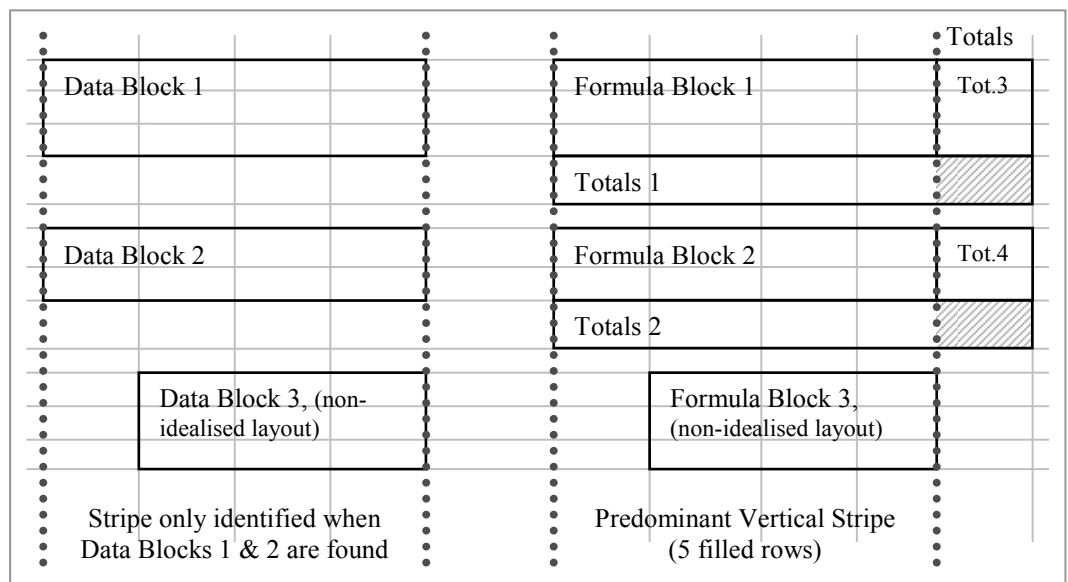

Figure 6: Finding a Predominant Vertical Stripe

### 5.1 Primary Passes (Idealised)

The first passes of the algorithm find those Blocks which obey the Idealised Layout Criteria, i.e. they would, on their own, meet CRIT.

5.1.1 Finding the Vertical and Horizontal "Stripes"

Figure 6 shows how a predominant Vertical Stripe is identified. This is achieved by invoking the "Find Formula Areas (Simple Find)" function (section 4.1 & 0) without skipping over "Exceptions". This function builds a temporary version of the "Map of Stripes" which contains all the different fill widths and heights encountered. The "Stripe Analysis" function (section 4.3) is then invoked to establish which stripes are predominant and contribute towards the structure of each sheet.

Considering figure 6 in more detail the two grey diagonal hatched cells on the right are column totals and are part of the filled Blocks of Totals 1 and 2. However, they need to be considered as separate because the row total formulas above them are different. The two filled Totals rows (1 & 2) could indicate a 5 column Stripe, but there are five rows (in total) in Formula Blocks 1 & 2 filled to 4 columns wide. This means that the 4 column Stripe is predominant in terms of the sheet structure. Another predominant 1 column Stripe (containing five formula rows (in total)) is also found formed by the Totals column, but this is of less importance as there is no horizontal fill.

The predominant Stripes form the layout structure of the sheet and are thus logged as conforming to an Idealised Layout. Formula Block 3 contains 3 filled rows and its 3 column Stripe overlaps with the previously defined predominant Stripe. This 3 column Stripe is thus logged as not conforming to an Idealised Layout.

At this early stage in the algorithm the stripe on the left of figure 6 has not been found as only the Formula Blocks have been examined.

5.1.2    Finding the Main Formulas (obeying Idealised Layout Criteria)

Now the predominant Stripes have been found the "Find Formula Areas (Complex Find)" function (section 4.1 & 4.1.2) is invoked finding filled Blocks, but skipping over "Exceptions" due to any corrupted formulas within filled areas. Only Blocks of filled formulas that match Stripes that conform to an Idealised Layout are processed at this stage.

In figure 6 Formula Blocks 1 & 2 are found together with the first 4 columns of Totals 1 & 2. Four Formula Blocks in the final totals column are found, including Totals 3 & 4 and the last cell of Totals 1 & 2 (diagonally shaded in grey).

5.1.3    Finding the Referenced Data (obeying Idealised Layout Criteria)

Having identified some formula Blocks that conform to an Idealised Layout the Find Data Areas function (section 4.2) is invoked looking at all references in these formulas. References that point to other Formula Blocks already found (e.g. the Totals column) are skipped but those pointing to cells which do not contain formulas are used to identify Data Blocks/Cells.

In figure 6 Data Blocks 1 & 2 are found from the references in Formula Blocks 1 & 2. The Stripe occupied by Data Blocks 1 & 2 is also identified and logged.

**5.2    Secondary Passes (Non-Idealised)**

Subsequent passes of the algorithm find those Blocks which do not obey the Idealised Layout Criteria i.e. they would not meet CRIT.

5.2.1    Re-Finding the "Stripes"

At this stage a second invocation of the "Find Formula Areas (Simple Find)" function (section 4.1 & 0) ensures that any smaller Blocks, within the larger blocks already found, are removed from the "Map of Stripes". The "Stripe Analysis" function (section 4.3) is then invoked, for a second time, to re-establish which stripes are predominant and refine the layout picture of each sheet. This will also assess the predominance of Stripes logged from Data Blocks which have already been found (e.g. the left hand predominant Stripe in figure 6).

5.2.2    Finding the Remaining Formulas (not obeying Idealised Layout Criteria)

Now the current set of predominant Stripes have been found the full "Find Formula Areas (Complex Find)" function (section 4.1 & 4.1.2) is invoked, for a second time, finding any remaining filled Blocks. All Blocks of filled formulas within all Stripes are processed at this stage, whether or not the stripes conform to an Idealised Layout.

In figure 6 Formula Block 3 is found as this is the only one left over because it did not conform to an Idealised Layout.

### 5.2.3  Finding the Remaining Referenced Data (not obeying Idealised Layout Criteria)

Having identified all the formula Blocks, including those that do not conform to an Idealised Layout, the Find Data Areas function (section 4.2) is invoked again, looking at all references in all formulas. References that point to other Formula and Data Blocks already found (e.g. the Formula Blocks 1 & 2) are skipped but those pointing to cells not found so far, and which do not contain formulas, identify additional Data Blocks/Cells.

In figure 6 Data Blocks 3 is found from the reference in Formula Block 3.

### 5.2.4  Data "Mop-Up" of any other Referenced Data

The final stage of the algorithm is to find any other data that is referenced but has been missed by all the stages so far. This will normally be data that is range referenced but which is outside or in between filled Formula Blocks, or Data Blocks that are referenced by filled Formula Blocks.

There are no data blocks of this type in figure 6 but figure 5 in section 4.2 shows the Unintentional Data Blocks 3 & 4 which would be found at this stage. These unintentional Blocks, in between other Blocks which are summed together, are important as any temporary data or sum formulas entered in these cells will affect the final totals.

## 6  CONCLUSION

The algorithms documented here have been implemented within our *F*ormula*D*ata*S*leuth[®] toolset and have successfully analysed a variety of different types of workbook from varying sources. The multi-pass nature of the algorithms allows the predominant structures to be identified first. This helps to identify Formula and Data Blocks that may be problematic in terms of the spreadsheet layout, particularly when modifications are made.

The main purpose of the Workbook Structure Analysis is to allow *F*ormula*D*ata*S*leuth[®] to manage an existing workbook and monitor it for damage. However, setting up the monitoring of a workbook requires the detail of the workbook structure to be defined and all the Formula and Data Blocks/Cells have to be identified. The procedure for doing this analysis has the spin-off of providing a comprehensive check on the structure of the sheets and how formula references access other related Blocks. It is possible to envisage the functionality described here as a separate entity from the error checking capabilities of our  *F*ormula*D*ata*S*leuth[®] toolset.